# Are Macula or Optic Nerve Head Structures better at Diagnosing Glaucoma? An Answer using AI and Wide-Field Optical Coherence Tomography

**Charis Y.N. Chiang**[1], **Fabian Braeu**[2,3,4], **Thanadet Chuangsuwanich**[1,2], **Royston K.Y. Tan**[1,2], **Jacqueline Chua**[5,6,7], **Leopold Schmetterer**[5,6,7,8,9,10,11], **Alexandre Thiery**[12], **Martin Buist**[1], **Michaël J.A. Girard**[2,7,13]

1. Department of Biomedical Engineering, National University of Singapore, Singapore
2. Ophthalmic Engineering & Innovation Laboratory, Singapore Eye Research Institute, Singapore National Eye Centre, Singapore
3. Singapore-MIT Alliance for Research and Technology, Singapore
4. Yong Loo Lin School of Medicine, National University of Singapore, Singapore
5. Singapore Eye Research Institute, Singapore National Eye Centre, Singapore
6. SERI-NTU Advanced Ocular Engineering (STANCE), Singapore, Singapore
7. Duke-NUS Graduate Medical School, Singapore
8. School of Chemical and Biological Engineering, Nanyang Technological University Singapore
9. Department of Clinical Pharmacology, Medical University of Vienna, Austria
10. Center for Medical Physics and Biomedical Engineering, Medical University of Vienna, Austria
11. Institute of Molecular and Clinical Ophthalmology, Basel, Switzerland
12. Department of Statistics and Data Sciences, National University of Singapore, Singapore
13. Institute for Molecular and Clinical Ophthalmology, Basel, Switzerland

| | |
|---|---|
| **Corresponding Author:** | Michaël J.A. Girard |
| | Ophthalmic Engineering & Innovation Laboratory (OEIL) |
| | Singapore Eye Research Institute (SERI) |
| | The Academia, 20 College Road |
| | Discovery Tower Level 6, |
| | Singapore 169856 |
| | mgirard@ophthalmic.engineering |
| | https://www.ophthalmic.engineering |

# Abstract

**Purpose**: **(1)** To develop a deep learning algorithm to automatically segment connective and neural tissue structures of the optic nerve head (ONH) and macula in 3D wide-field optical coherence tomography (OCT) scans; **(2)** To exploit such information, to assess whether 3D macula or ONH structures (or the combination of both) provide the best diagnostic power for glaucoma.

**Methods:** A cross-sectional comparative study was performed which included wide-field swept-source OCT scans from 319 subjects with primary open angle glaucoma (POAG) and 298 non-glaucoma subjects. After acquisition, all scans were compensated to improve deep-tissue visibility. We developed a deep learning algorithm to automatically label all major neural and connective tissue structures by using 270 manually annotated B-scans (from 69 eyes with POAG and 48 eyes without POAG) for training. The performance of our algorithm was assessed using the Dice coefficient (DC). A classification algorithm (3D-Convolutional Neural Network) was designed using a combination of 500 OCT volumes and their corresponding automatically segmented masks to classify the scans as non-POAG or POAG. This algorithm was trained and tested on three similar datasets: OCT scans cropped to contain the macular tissues only, those cropped to contain the ONH tissues only, and the full wide-field OCT scans containing both regions. The classification performance for each dataset was reported using the area under the receiver operating characteristic curve (AUC) score and their diagnostic capabilities compared.

**Results:** Our segmentation algorithm was able to segment major ONH and macular tissues with a DC of 0.94 ± 0.003. The classification algorithm was best able to diagnose POAG using wide-field 3D-OCT volumes with an AUC of 0.99 ± 0.01, followed by ONH volumes with an AUC of 0.93 ± 0.06, and finally macular volumes with an AUC of 0.91 ± 0.11.

**Conclusions:** The AI-based segmentation method developed here was able to identify major tissue layers in the wide-field OCT scans with high levels of accuracy. This study also suggests that wide-field OCTs lead to the best classification of POAG and non-POAG as compared to ONH and macular OCT scans, while OCT scans of the ONH are superior to macular OCT scans for POAG and non-POAG classification. Such results could form a basis for the mainstream adoption of 3D-OCT wide-field scans for POAG diagnosis.

# Introduction

Primary open-angle glaucoma (POAG) is one of the world's leading causes of permanent blindness [1], [2]. POAG is a progressive and chronic form of glaucoma [3], characterized by distinct damage to the optic nerve [4], [5]. This damage includes the excavation, or cupping, of the ONH and progressive loss of retinal ganglion cells (RGC) and RGC axons in the ONH and macular, leading to vision loss [6], [7]. Up to 50% of one's total RGCs could potentially be destroyed before any noticeable symptoms of vision loss [8], thus, early detection is crucial to stop disease progression and prevent irreversible damage [9].

Due to the nature of glaucomatous damage, especially in earlier stages of the disease, it is crucial to identify three-dimensional (3D) structural changes in both the macular and ONH. Thus, the use optical coherence tomography (OCT) to obtain 3D images of either the macular or ONH has proven valuable for diagnosing and monitoring POAG [10]. Recently, a new form of wide-field 3D-OCT has emerged which can image both the ONH and macular in one scan. These scans could be useful to detect more subtle structural changes, which could prove crucial for earlier detection.

Diagnostic studies using OCT in glaucoma have focused on either macular or ONH volumes. For example, Asaoka et. al. applied various AI methods to diagnose POAG from 3D macular OCT volumes. They used 3D-Convolutional Neural Network (CNN), Random Forest and Support Vector Machine methods and achieved an AUC of 0.937 with their 6-layer 3D-CNN model [11]. Similarly, George et. al. used a 3D-CNN AI method to diagnose POAG from a large datasets of raw 3D OCT volumes centered on the ONH. Their 8-layer 3D-CNN approach achieved an AUC of 0.973 for glaucoma detection [12]. Non-AI methods have also been employed for glaucoma diagnosis, including a study from Mori et. al. that achieved an AUC of 0.922 using structural measurements derived from macular scans [13]. With wide-field OCT devices, it is now possible to obtain 3D volumes that contain both the ONH and macular in a single scan. Access to this wider area allows us to explore and compare the utility of using each separate structure along with the combination of both to diagnose glaucoma [14].

In this study we developed a deep-learning approach to automatically and simultaneously identify seven major neural and connective tissue structures of the ONH and macular region from 3D wide-field OCT scans. We then exploited this information using a subsequent AI classification algorithm to assess whether 3D macula or ONH structures (or the combination of both) provide the best diagnostic power for glaucoma.

## Methods

### Patient Recruitment

A total of 617 subjects (298 non-POAG subjects and 319 subjects diagnosed with POAG) were retrospectively included in this study at the Singapore Eye Research Institute (SERI, Singapore). All subjects gave written informed consent. The study adhered to the tenets of the Declaration of Helsinki and was approved by the institutional review board of the institution. POAG was diagnosed if the subject met the following criteria: (1) presence of glaucomatous optic neuropathy, defined as vertical cup-to-disc ratio of > 0.7 or inter-eye asymmetry of > 0.2 and/or notching that can be attributed to glaucoma, (2) compatible and reproducible visual fields in standard automated perimetry, (3) a glaucoma hemifield test outside normal limits, with mean deviation better than -6dB, reflected open angles on gonioscopy, and (4) did not have secondary causes of glaucomatous optic neuropathy [15]. The POAG subjects were using the following eye-drops to lower intra-ocular pressure: beta blockers (timolol), alpha-2 adrenergic agonists (brimonidine), prostaglandin analogs (latanoprost, bimatoprost, travoprost, tafluprost), and carbonic anhydrase inhibitors (brinzolamide). Individuals classified as non-POAG were those not diagnosed with any clinically relevant eye conditions, including glaucoma, age-related macular degeneration, diabetic retinopathy, and ocular vascular occlusive disorders, diabetes and other causes of neuro-ophthalmic disease [15].

**Optical Coherence Tomography Imaging**

OCT imaging was performed on seated subjects in a dark room, and tropicamide 1% solution was used when pupil dilation was necessary. In total 617 3D-OCT wide-field volumes were taken, 319 from subjects diagnosed with POAG and 298 from non-POAG subjects. The swept-source OCT machine (PlexElite 9000, Zeiss Meditec, Dublin, CA, US), operating at 1060 nm, was used to take 12mm x 12mm wide-field 3D-OCT scans. All OCT volumes (horizontal raster scans) covered the entire ONH and macula with 500 B-scans (slices), and 1,536 A-scans per B-scan. The number of pixels per A-scan was 500. In terms of resolution, the distance between B-scans was 24 µm, the lateral resolution was 20 µm, and the axial resolution was 1.95 µm.

**Correction of Light Attenuation Using Adaptive Compensation**

To remove the deleterious effects of light attenuation from OCT images, all B-scans were post-processed using adaptive compensation [16], [17]. For OCT images of the ONH, adaptive compensation has been shown to mitigate blood vessel shadows, improve tissue contrast, and increase the visibility of tissue layer boundaries, especially for deep tissues such as the lamina cribrosa (LC) [18]. For all B-scans, we used a decompression and compression exponent of four to limit noise over-amplification, and a contrast exponent of two to improve overall image contrast. This step was critical to visualize deep ONH connective tissues before performing manual segmentation.

**Manual Segmentation of Seven Tissue Layers in OCT Images**

To train a deep learning algorithm to automatically label all major neural and connective tissues in a wide-field OCT scan, we manually segmented 270 OCT images (B-Scans), including 135 B-scans taken from 48 non-POAG eyes and 135 B-scans taken from 69 POAG eyes. These were randomly sub-sampled from a larger dataset of wide-field OCT scans. Each compensated OCT image was manually segmented using Amira (version 6, FEI, Hillsboro, OR) to identify the following tissue groups as shown in Figure 2: **(1)** the RNFL and the prelaminar tissue (PLT); **(2)** the ganglion cell layer (GCL) and the inner plexiform layer (IPL); **(3)** the retinal pigment epithelium (RPE); **(4)** all other retinal layers; **(5)** the choroid; **(6)** the peripapillary and posterior sclera; and **(7)** the LC. It should be noted that in most cases (especially POAG subjects), we could not achieve full-thickness segmentation of the peripapillary sclera and of the LC due to limited visibility at high depth, even when compensation was used [19]. Therefore, only the OCT-visible portions of the sclera and LC were segmented as per the observed compensated signal, and no efforts were made to capture their true thickness. The manual segmentation assigned a label to each pixel of each OCT image to indicate the tissue type as described above, with the background assigned a value of zero.

**Automated Segmentation of Tissue Layers using Deep Learning**

To automatically segment all tissue layers of the ONH and macula, we used a Unet++ model with ResNet-34 encoder as the backbone, implemented in Pytorch. Unet++ is a nested U-net architecture designed to improve the accuracy and efficiency of medical picture segmentation operations [20]. This is achieved via adding convolutional layers on skip pathways and dense blocks, inspired by DenseNet, between the encoder and decoder of the original U-net architecture and using deep supervision to allow for pruning of the model [20].

The encoder extracted features from the input images that the decoder used to create segmentation masks while the skip connections were utilized to prevent the vanishing gradient problem and help in the backpropagation process. For the encoder, we utilized the ResNet-34 network with pre-trained weights (on ImageNet), one of the most cutting-edge networks that makes use of residual connection [21]. We used the Jaccard index (mean for all tissues) to represent the loss function. To avoid overfitting during the training process, we performed extensive data augmentation using the python library Albumentations. Image transformations such as horizontal flipping, random rotation and translation, additive Gaussian noise, and random saturations were used as data augmentation to enrich the training set. These transformations were shown to significantly improve performance of the segmentation model.

Images from the segmentation dataset were split into training (192 B-scans), validation (48 B-scans), and test (30 B-scans) sets, respectively. The split was performed in such a way that the same image that had been augmented or images from the same subject did not exist in different sets. We also ensured that half of the images in each set were of POAG eyes and the other half were of non-POAG eyes. All B-scan images were resized to 320 x 480 pixels. Our network was then trained on an Nvidia 1080Ti GPU card until optimum performance was reached in the validation set which was about 3,000 epochs (computational time: ~48 hours). To evaluate the accuracy of our segmentation network, we calculated the Dice coefficient by comparing the network predicted labels with the corresponding manually segmented images from the test set. Dice coefficients are reported as mean ± standard deviation through a 5-fold cross-validation process.

**Wide-Field OCT scans divided into ONH and Macular Regions**

To determine whether the use of wide-field scans improved glaucoma diagnosis over existing methods, each of the wide-field OCT scans was divided into two sections: one including all ONH tissues, and another the macular tissues, as shown in Figure 3. The partition was done such that the ONH portion represented one quarter of the total scan width, and the macula section three quarters, which ensured that the entire ONH was included and approximately centered for all ONH portions. To reduce computational time for AI training, the scans were all down-sampled by the same factors, resulting in three similar datasets with the following sizes: (1) 500 whole wide-field volumes (167 slices with 125 x 192 pixels); (2) 500 ONH volumes (167 slices with 125 x 48 pixels); and (3) 500 macular volumes (167 slices with 125 x 144 pixels).

**Diagnostic Performance Comparison using ONH, Macula or Wide-Field Scans**

Binary classification was used to identify whether a given 3D-OCT volume (Macula, ONH, or whole wide-field) was classified as either 'non-glaucoma' or 'glaucoma'. For such classifications, we opted to use a 3D-CNN. It was necessary to approach the classification problem with a dataset distinct from the one used to train our segmentation network to prevent bias for the two independent tasks. We used three volumes (macular, ONH and whole wide-field) for classification. Each dataset consisted of 250 POAG and 250 non-POAG 3D-OCT image volumes, which were the remaining volumes not used in the segmentation model. The 3D image volumes were compensated and segmented tissue labels were also generated using the segmentation model trained earlier. To increase the sample size, data augmentation was performed with Albumentations using horizontal flips and slight rotations and translations. Each of the three training datasets were then split into 80% training plus validation dataset and 20% test datasets.

Subsequently, three classification models with identical architecture were trained on each of the training datasets. The datasets were used as an input to the 3D CNN model to classify a segmented OCT volume into one of two classes: 0 (POAG) or 1 (non-POAG). The network architecture was built using the TensorFlow Keras library. The two network inputs, being compensated 3D-widefield volume data and corresponding segmentation masks, were each separately fed into two 3D-convolutional blocks. Next, the results of each convolutional blocks were concatenated to produce a single classification output. This concatenation was done to allow the model to learn information from both the scans and the segmentations,

Each of the networks was then tested on the test dataset. We performed five-fold cross validation and reported the area under the receiver operating characteristic curves (AUCs; mean and standard deviation) for the classification performance on each of the test datasets for the three cases. The mean AUC of using either the macula or ONH section or both were compared. An overview of the methodology from acquired raw OCT scans to final POAG classification is shown in Figure 1.

# Results

**Wide-field Segmentation**

Compensated, manually segmented and AI-segmented images from the test dataset are shown in Figure 4. Our network was able to identify several major neural and connective tissues in a wide-field OCT scan with a DC of 0.94 ± 0.003, a Jaccard loss of 0.11 ± 0.02. Discrepancies were noted in some of the automated-segmentation results, mainly: **(1)** LC with an inconsistent thickness throughout its width; **(2)** mismatched LC boundaries in automatically segmented images as compared to manual segmentations; and **(3)** noise segmented as tissue layers (large islands), this noise occurred in less than 5% of the test dataset.

**Classification performance using different anatomical regions**

An AUC of 0.91 ± 0.11 was achieved when using the macula portion of the scan to classify POAG and non-POAG scans. With the ONH portion of the scan, the resultant AUC of 0.93 ± 0.06 was better for the same classification. Finally, classification of POAG and non-POAG whole wide-field scans of both the macular and ONH resulted in the best AUC of 0.99 ± 0.01. An example of the classification of four separate 3D-OCT volumes is shown in Figure 5 with one B-scan from each of the scan volumes displayed.

**Misclassification of Wide-Field OCT scans: Qualitative Analysis**

As seen in the example as shown in Figure 5, scans that were wrongly classified typically had segmented noise or segmentation errors especially around the LC area. This trend was seen in each of the models and highlights the importance of an accurate segmentation to prevent segmentation errors contributing to glaucoma misclassification.

# Discussion

In this study, we explored whether the new wide-field 3D OCT scan technology allowed for improved POAG diagnosis as opposed to 3D OCT scans of the ONH or macular region alone. This was done via two steps. First, the proposed deep learning algorithm was able to accurately segment tissue layers of the ONH and macular in the wide-field scans. Second, by exploiting the information from both the scan and segmentations we were able to classify POAG and non-POAG scans via three similar 3D CNN models. We found that the classification performance of the wide-field OCT scans surpassed both of the traditional scan types, while the diagnostic ability of the ONH OCT scans was superior to that of the macular OCT scans.

This study showed that it was possible to segment 3D-OCT wide-field scans accurately. The segmentation algorithm was able to simultaneously isolate seven classes of tissue layers, achieving very good segmentation performance with Dice coefficient of 0.94 ± 0.003. Although this may be the first study segmenting wide-field scans, the resulting performance is comparable to the top algorithms for segmenting regular OCT scans of the ONH, as reviewed by Marques et. al [22]. This includes Devalla et. al's models from 2018 and 2022 which achieved mean $\pm$ SD dice coefficients of 0.91 $\pm$ 0.05 and 0.93 $\pm$ 0.02, respectively, and Yu et. al's 2018 model which achieved mean $\pm$ SD Dice coefficients of 0.925 ± 0.03 [23]–[25]. Furthermore, our model segmented more than or as many key tissue layers as the top performing algorithms [22]–[26]. Furthermore, although there were some errors with segmentation, especially in the LC layer, these were also common errors reported in the aforementioned automated segmentation models [22]–[24], [26] .

Using a 3D CNN, we found that the diagnostic performance of the ONH scans, which achieved an AUC of 0.93 $\pm$ 0.06, was better than that of the macular scans, which achieved an AUC of 0.91 $\pm$ 0.11. This suggests that the ONH region in wide-field scans provides more information than the macular region for POAG and non-POAG classification. This result was echoed in Wollstein *et al.*'s study that found that using ONH OCT scans to discriminate between glaucoma and non-glaucoma scans was better than using macular OCT scans [27]. One possible reason for this result could be that ONH scans capture 100% of RGCs whereas macular scans capture only 50% of RGCs, thus allowing ONH scans to provide better discrimination of POAG than macular scans [28]. Another factor could be that connective tissues, and especially the LC, which undergoes remodeling with loss of curvature and posterior bowing caused by POAG, is only visible in the ONH scans [29]–[31]. Other studies that used 3D OCT scans of the ONH for deep learning-based glaucoma classification also generally achieved higher AUC scores than studies which used 3D OCT scans of the macular [32]. These results indicate that ONH OCT scans, instead of macular OCT scans, should be preferred for glaucoma studies employing AI or deep learning where wide-field scans are not available.

The results of the 3D CNN classification further revealed that the diagnostic performance of the wide-field scans surpassed both the macular and ONH scans by achieving $0.99 \pm 0.01$ AUC, corresponding to an increasingly accurate classification of POAG. This suggests that the additional information provided by including both the ONH and macula, as reflected in wide-field scans, adds value to the POAG classification process. The 3D CNN with wide-field scans also performed slightly better than other similar POAG diagnosis methods using deep learning applied to macular or ONH scans from the literature. This includes various 2D CNN methods performed on fundus images of the ONH which achieved AUCs between 0.80 and 0.90, 3D CNN applied to unsegmented 3D OCT scans of the ONH which resulted in an AUC of 0.973, and 3D CNN performed on unsegmented macular 3D OCT scans which achieved an AUC of 0.937 [11], [12], [33]. It should be noted that a direct comparison of AUC across different studies may not be entirely accurate due to different datasets and experimental conditions being employed. However, the higher AUC our wide-field scans achieved as compared to our own macular and ONH datasets as well as others from the literature does suggest that wide-fields could allow for more accurate diagnosis of POAG.

Some limitations in this study warrant further discussion. First, there could be errors in the manual segmentation dataset as parts of the sclera and LC were not fully visible in some of the scan volumes and were segmented by an educated guess. This could perhaps explain why the segmentation model did not perform as well in segmenting the LC tissue layer. Going forward, each manual segmentation could be performed by more than one individual, thus mitigating observer bias.

Second, there was a propagation of errors from the automated segmentation to the classification model. Though these errors were rare, they likely contributed to misclassification. Possible mitigations include AI-based post-processing to identify and correct these errors, or to develop improved automated segmentation models, taking the mean, mode or weighted mean of the results as the final resultant class label for each pixel in the segmentation.

Third, the OCT images were not corrected for optical distortions which often affect the posterior segment of the eye, including the ONH, and which could have affected the 3D analysis of the scans [34], [35]. These artificial deformations to the ONH tissue shapes could potentially be more exaggerated in the wide-field scans, hampering POAG classification, especially for early stage POAG where the tissue deformations caused by POAG are often smaller. Possible correction methods could include Grytz's correction equation derived for nonlinear distortion correction or Kuo's numeric correction approach derived from modelled linear scan beams [34], [35].

A fourth limitation of the study was the relatively small dataset sizes may lead to generalization errors. These dataset sizes for both the segmentation and classification model were comparable to that of other similar problems involving 3D-OCT scans because of the difficulties involved in obtaining a large set of manually segmented scans. However, when comparing with deep-learning best practices, the datasets are rather small and using a larger dataset may result in a more reliable and accurate assessment of the model's performance.

Finally, there were constraints in computational memory and processing time which resulted in a down-sampling of the training images. This could have contributed to a loss of information, especially for some of the thinner tissue layers such as the RNFL and RPE. Ideally the original resolution of the training data could be preserved.

In conclusion, this study showed that using wide-field OCT as compared to the typical OCT images containing just the ONH or macular may allow for a significantly improved POAG diagnosis. This may encourage the mainstream adoption of 3D-OCT wide-field scans. For clinical AI studies that use traditional machines without a wide-field capability, we would recommend the use of ONH scans as opposed to macula scans.

## Acknowledgments

Acknowledgement is made to **(1)** the donors of the National Glaucoma Research, a program of the BrightFocus Foundation, for support of this research (G2021010S [MJAG]), **(2)** NMRC-LCG grant 'TAckling & Reducing Glaucoma Blindness with Emerging Technologies (TARGET)', award ID: MOH-OFLCG21jun-0003 [MJAG], **(3)** SingHealth Duke-NUS Academic Medicine Research Grant (SRDUKAMR21A6 [MJAG]), and **(4)** the "Retinal Analytics through Machine learning aiding Physics (RAMP)" project that is supported by the National Research

Foundation, Prime Minister's Office, Singapore under its Intra-Create Thematic Grant "Intersection Of Engineering And Health" - NRF2019-THE002-0006 awarded to the Singapore MIT Alliance for Research and Technology (SMART) Centre [MJAG].

# Figures

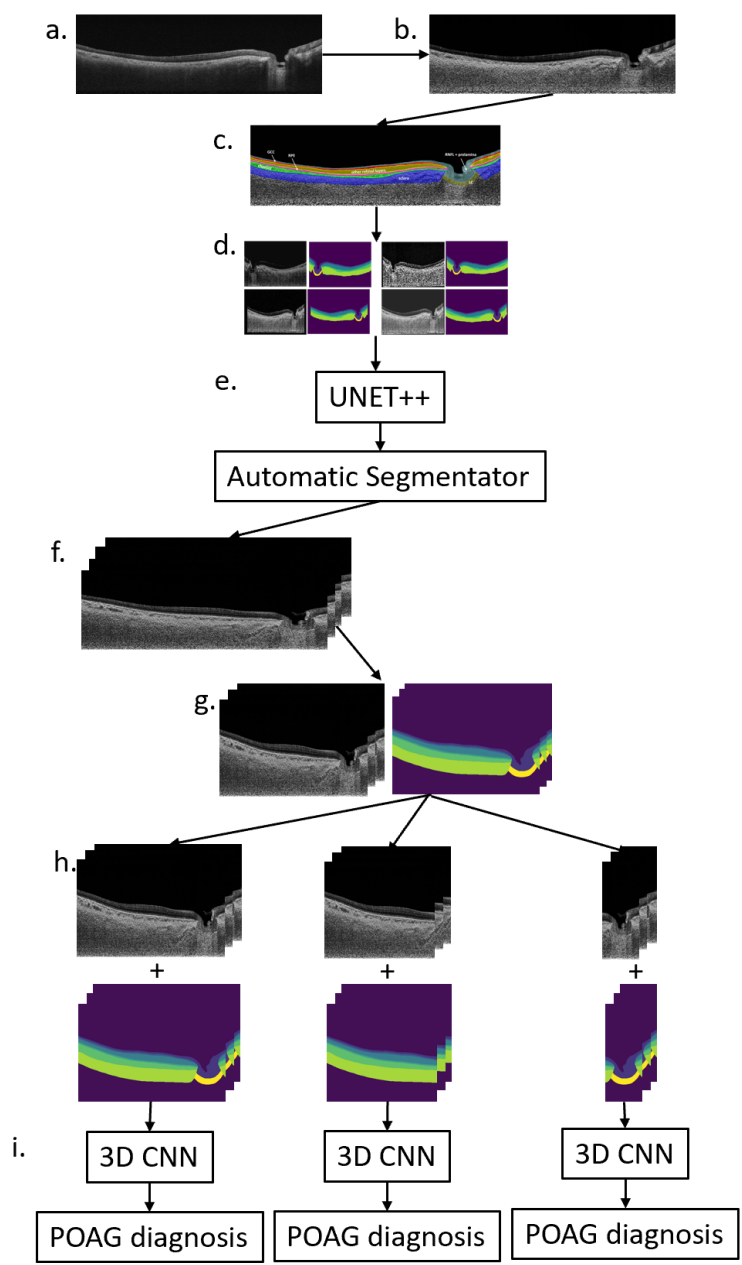

**Figure 1. a.** Raw slices (b-scans) of wide-field scans were **b.** compensated, **c.** manually segmented, **d.** down-sampled, augmented, **e.** and used to train a Unet++ segmentation model. **f.** Next, wide-field scan volumes were compensated, **g.** automatically segmented, down-sampled, **h.** and cropped into separate wide-field, macular and ONH regions. **i.** Each of these separate scan regions (both compensated volumes and automatic segmentations) were used to train separate similar 3D CNN and the results of each of these classifiers to diagnose POAG were compared.

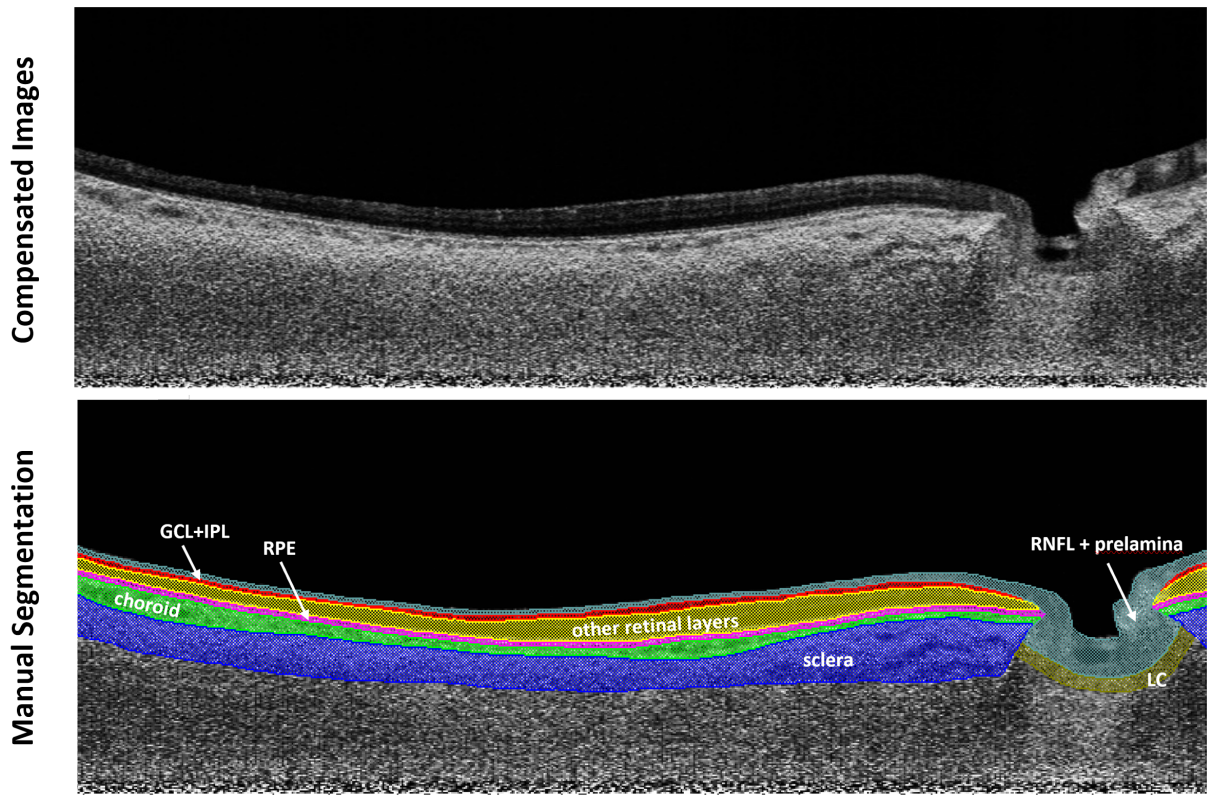

**Figure 2.** To train our AI algorithm, manual segmentation was performed on OCT wide-field images from both POAG and non-POAG subjects. The following tissues (or tissue groups) were identified: (1) RNFL + prelamina, (2) GPL + IPL, (3) other retinal layers, (4) RPE, (5) choroid, (6) sclera, and (7) lamina cribrosa. Baseline images (compensated) are shown 1st row while tissue boundaries are shown on the 2nd row.

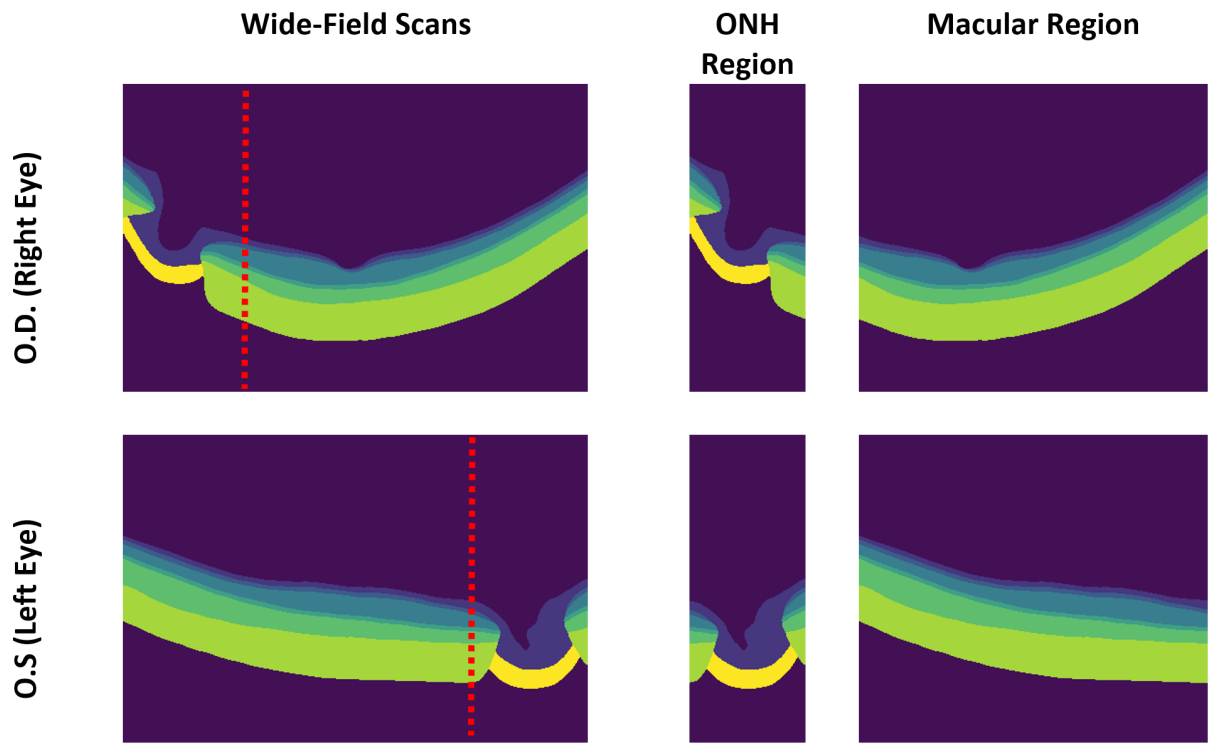

**Figure 3.** The wide-field scans (1st column) were cropped into ONH region (2nd column) and macular region (3rd column) at a fixed point along its width. Each full slice in the volume has a width of 192 pixels. Each slice in the scan is cropped along its width resulting in an ONH scan of 48 pixels width, and a macular scan of 144 pixels width. An example of a slice from a scan of the right and left eyes are shown in the 1st and 2nd row respectively.

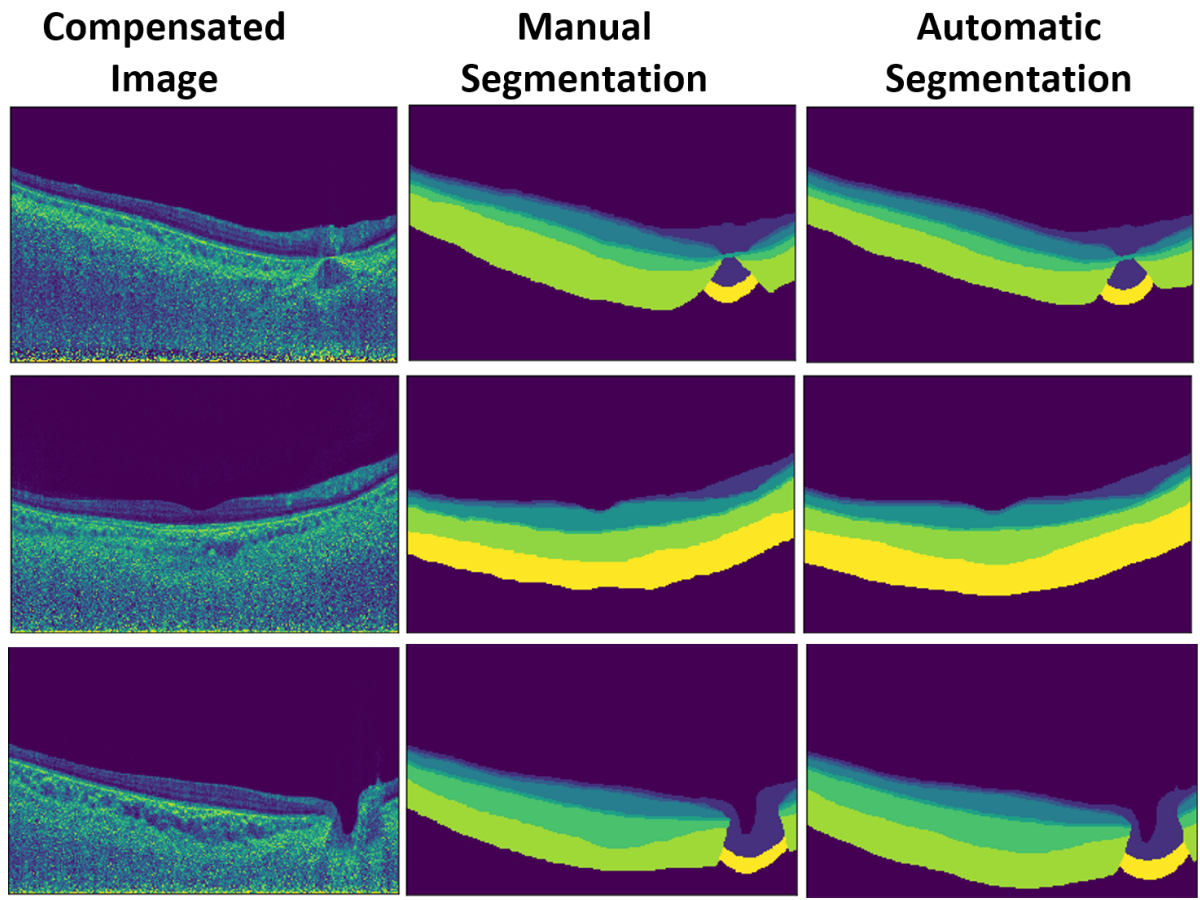

**Figure 4.** Three compensated images (slices), its corresponding manual segmentation (ground truth) and automatic segmentation by the algorithm. The algorithm was largely able to identify each of the tissue layer boundaries.

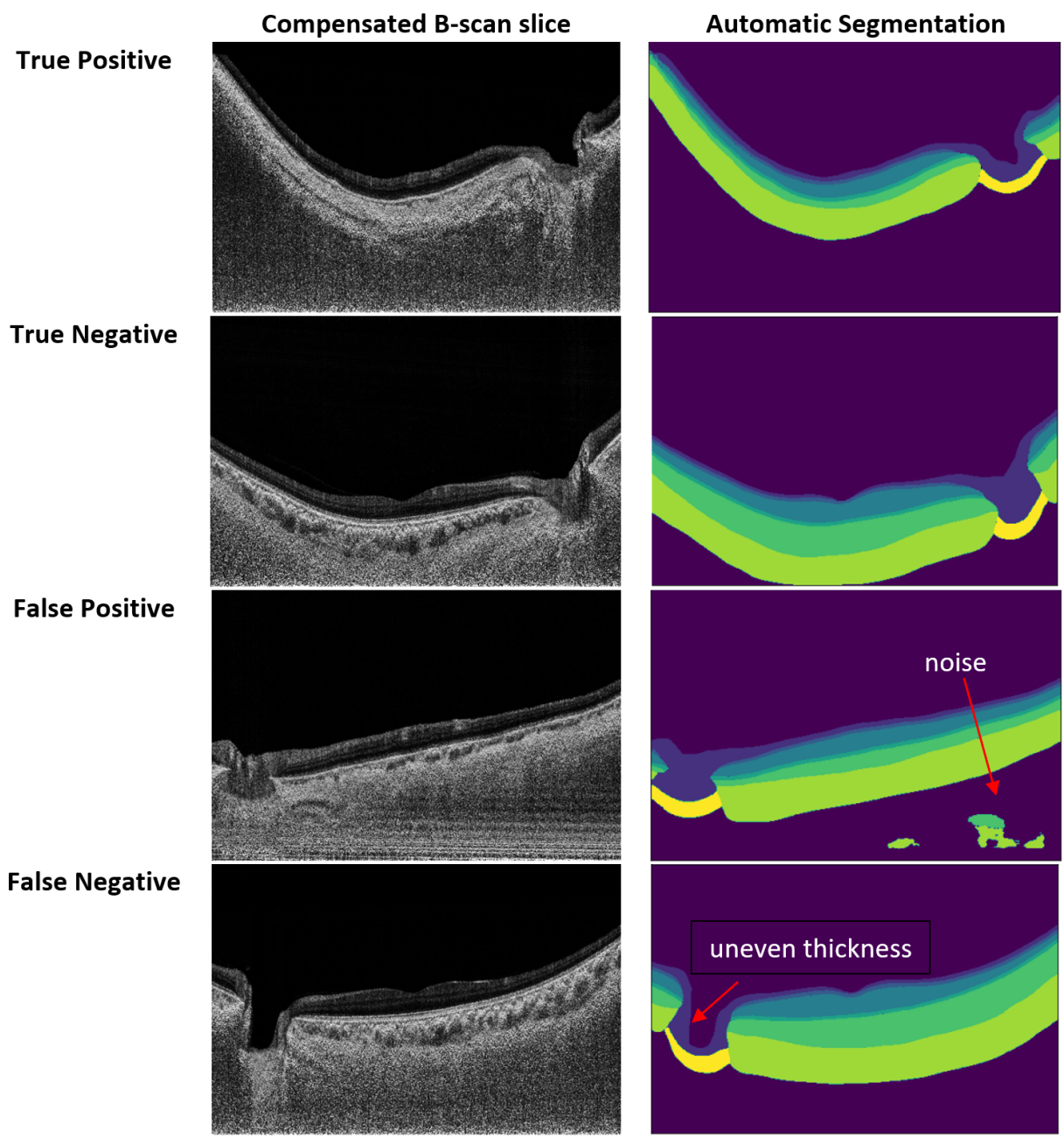

**Figure 5.** Slices from segmented scan volumes that were classified as: (1) a true positive, correctly classified as POAG; (2) a true negative, correctly classified as non-POAG; (3) a false negative, wrongly classified as non-POAG, and (4) a false positive, wrongly classified as POAG